\documentclass[journal]{igtesymp}

\usepackage{graphicx}
\usepackage{amsmath}
\usepackage{siunitx}
\usepackage{xspace}
\usepackage{ninecolors}
\usepackage[nolist]{acronym}
\usepackage{tudacolors}
\usepackage{csquotes}
\usepackage{placeins}

\usepackage{tikz}
\usepackage{pgfplots}
\usepackage{tikz-3dplot}
\usepackage{ifthen}
\usepackage{pgffor}
\usepackage{pgfmath}

\usetikzlibrary{decorations.pathmorphing, decorations.pathreplacing, decorations.shapes,decorations.markings,math,3d,circuits.ee.IEC,calc,arrows.meta,positioning, shapes,fit,calc,chains}
\pgfplotscreateplotcyclelist{myColorCycleList}{black,TUDa-2b,TUDa-9b,TUDa-3c,TUDa-7b,TUDa-11b,TUDa-10c,TUDa-8c}
\usetikzlibrary{pgfplots.colormaps,patterns, patterns.meta}
\usepgfplotslibrary{colorbrewer, patchplots}
\pgfplotscreateplotcyclelist{viridis few}{[samples of colormap=4 of viridis]}
\pgfplotscreateplotcyclelist{viridis many}{[samples of colormap=9 of viridis]}
\pgfplotsset{cycle list name={myColorCycleList}}

\pgfplotsset{compat=1.18}
\pgfplotsset{every axis plot/.style={thick,mark=none}}
\tikzset{>=Latex}
\pgfplotsset{field plot element/.style={line width=0pt,faceted color=none}}
\pgfplotsset{field plot node/.style={line width=0pt,faceted color=none,shader=interp}}
\usepackage[europeaninductors]{circuitikz}

\RequirePackage{fontawesome}

\usepackage[style=numeric-comp,maxnames=3, minnames=1,nohashothers=true,isbn=false,doi=false,url=false,sorting=none]{biblatex}

\usepackage[colorlinks=false, hidelinks]{hyperref}
\usepackage{tikz}
\usepackage{pgfplots}
\usepackage[english]{babel}
\usepackage{bm}

\pgfplotsset{compat=newest}
\pgfplotsset{plot coordinates/math parser=false}
\newlength\figureheight
\newlength\figurewidth
\newlength\fwidth

\title{Thermal Finite Element Modeling and Simulation of a Squirrel-Cage Induction Machine}

\author{Christian Bergfried\IEEEauthorrefmark{1}\textsuperscript{,}\IEEEauthorrefmark{2}, Yvonne Späck-Leigsnering\IEEEauthorrefmark{1}\textsuperscript{,}\IEEEauthorrefmark{2}, Roland Seebacher\IEEEauthorrefmark{3}, Heinrich Eickhoff\IEEEauthorrefmark{4} and Annette Muetze\IEEEauthorrefmark{3}\\
	\vspace{0.3cm} \normalsize{
		\IEEEauthorblockA{\IEEEauthorrefmark{1}Institute for Accelerator Science and Electromagnetic Fields (TEMF), TU Darmstadt, Germany\\
			\IEEEauthorrefmark{2}Graduate School of Excellence Computational Engineering, 64293 Darmstadt, Germany\\
			\IEEEauthorrefmark{3}Electric Drives and Machines Institute (EAM), TU Graz, Austria\\
			\IEEEauthorrefmark{4}Silicon Austria Labs GmbH, Graz, Austria\\
			E-mail: \href{mailto:christian.bergfried@tu-darmstadt.de}{christian.bergfried@tu-darmstadt.de}}}
}

\IEEEaftertitletext{\vspace{-1cm}\noindent%
	\begin{abstract}%
		 Finite element models of electrical machines allow insights in electrothermal stresses which endanger the insulation system of the machine. This paper presents a thermal finite element model of a 3.7 kW squirrel-cage induction machine. The model resolves the conductors and the surrounding insulation materials in the stator slots. A set of transient thermal scenarios is defined and measured in the machine laboratory. These data are used to assess the finite element model.
	\end{abstract}%
	\noindent%
	\vspace{\baselineskip}
}

\begin{document}
\maketitle

\section{Introduction}\label{sec:introduction}
The insulation system of an electrical machine is a complex arrangement of different insulating materials \cite{Chapman_2008aa}. Its thermal and mechanical stability under adversary operating conditions defines the permissible operating range and determines the lifetime of a machine \cite{Pyrhonen_2013aa, CigreA117_2013aa}. Currently, the reliability and lifetime of the insulation system during transient thermal loads are usually not considered during the design process \cite{Madonna_2020aa}. As a consequence, many machines are overdimensioned, whereas others are operated beyond their thermal tolerances \cite{Nussbaumer_2015aa}. A detailed understanding of the thermal stresses inside the insulation system is, thus, of great importance for the improvement of the cost-effectiveness and robustness of electrical machines. 

Experimental investigations and diagnostic approaches for machine insulation systems are limited \cite{Montanari_2002aa,Elspass_2022}. Measurement campaigns are expensive, cumbersome and slow down the design process. A promising alternative are field simulations that are able to adequately resolve the thermal stresses inside insulation systems in space and time. 
This work presents a complete thermal simulation model of a 3.7\,kW squirrel-cage induction machine. 
The \ac{2d} \ac{fe} model is implemented in the open-source framework \textit{Pyrit} \cite{Bundschuh_2023ab}. Since the interturn insulation is considered to be the weakest part of the insulation system \cite{Driendl_2022aa}, the model explicitly resolves the arrangement of conductors and insulating components within the stator slots.
A series of thermal and electrothermal measurements of a 3.7\,kW induction machine has been conducted at the Electric Drives and Machines Laboratory of the TU Graz \cite{Eickhoff_2021aa} and the simulation model is successfully validated against the experimental data. 
The presented model allows the investigation of various modes of machine operation, such as on-off cycling, switch-on and operation at peak load, and thereby adds to a deeper understanding of thermal stresses inside the machine's insulation.  

\section{Induction Machine and Measurement System}
\begin{figure}
	\centering
	\includegraphics[width=\linewidth]{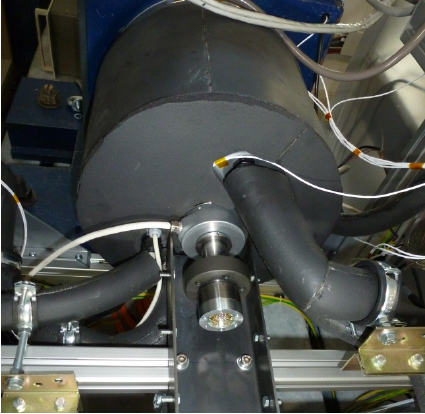}
	\caption{3.7\,kW squirrel-cage induction machine at the Electric Drives and Machines Laboratory of the TU Graz. A heat-insulating layer is wrapped around the cooling jacket of the machine. The figure shows the non-drive side with mounted rotary encoder and the shaft end rotor telemetry.}
	\label{fig:1}
\end{figure}

\begin{figure}
	\centering
	\includegraphics[width=.8\linewidth]{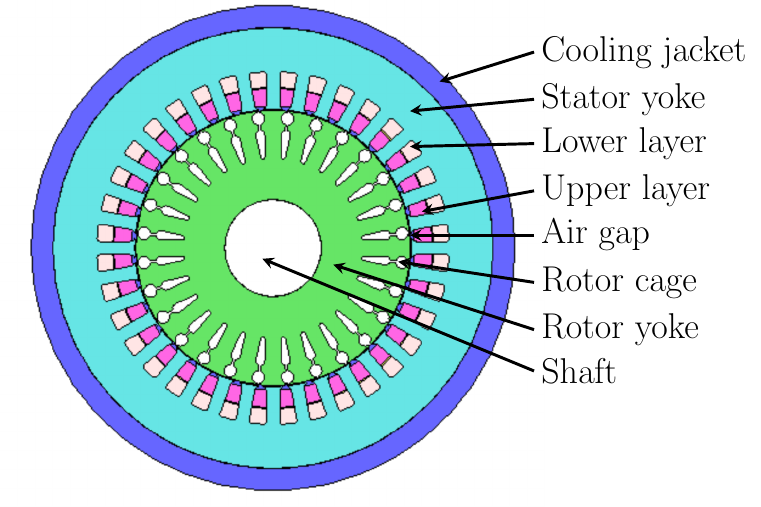}
	\caption{ Schematic of the \ac{2d} machine model. The machine shaft is located in the center of the rotor and colored in white. The rotor yoke is displayed in green. The squirrel cage is shown in yellow. The stator yoke is shown in light blue, and is surrounded by the cooling jacket (dark blue). The main area of interest in this work is the slot area filled with a dual layer copper winding that is insulated with epoxy resin. The upper and lower winding segments are displayed in two shades of pink. For clarity, the conductor and insulation system arrangement are not shown in this figure.}
	\label{fig:2}
\end{figure}

Figure~\ref{fig:1} shows the investigated four-pole, squirrel-cage induction machine. The machine is designed for a nominal power of 3.7\,kW at a nominal speed of 1430\,rpm. The stator has a distributed dual layer winding with three slots per pole and phase and a 7/9 pitching. There are 36 slots in the stator. Each winding is comprised of 18 conductors with a radius of 0.75\,mm. Further parameters are summarized in Tab.~\ref{tab:1}.
A schematic of the machine's cross section is shown in Fig.~\ref{fig:2}. 
The rotor is located at the center of the machine and encircled by the stator. The rotor consists of a steel shaft in the middle, a laminated iron yoke and an aluminium cage. The stator yoke also consists of laminated iron and of slots filled with a dual layer copper winding which are insulated with epoxy resin.

\begin{figure}
	\centering
	\includegraphics[width=0.8\linewidth]{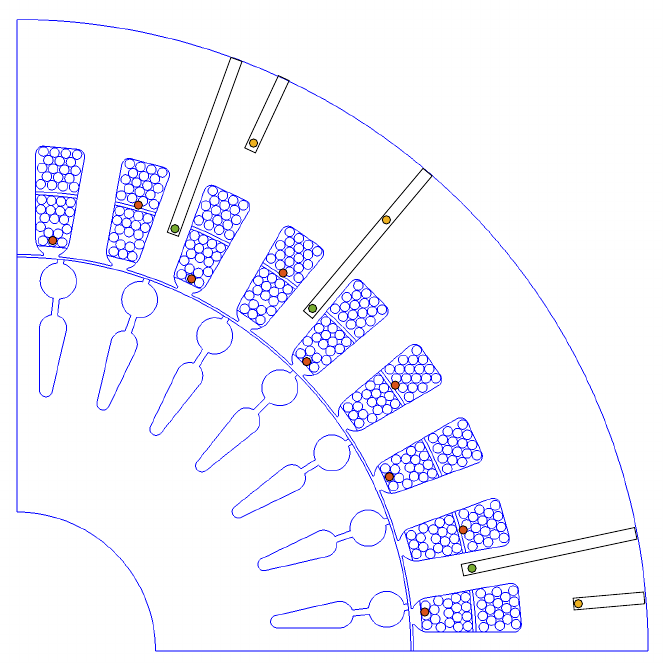}
	\caption{Positions of the temperature sensors inside the stator at the axial center. Sensors in the stator slot are red, sensors in the stator teeth are green and sensors in the stator yoke are yellow.}
	\label{fig:3}
\end{figure}

\begin{figure}
	\centering
	\includegraphics[width=\linewidth]{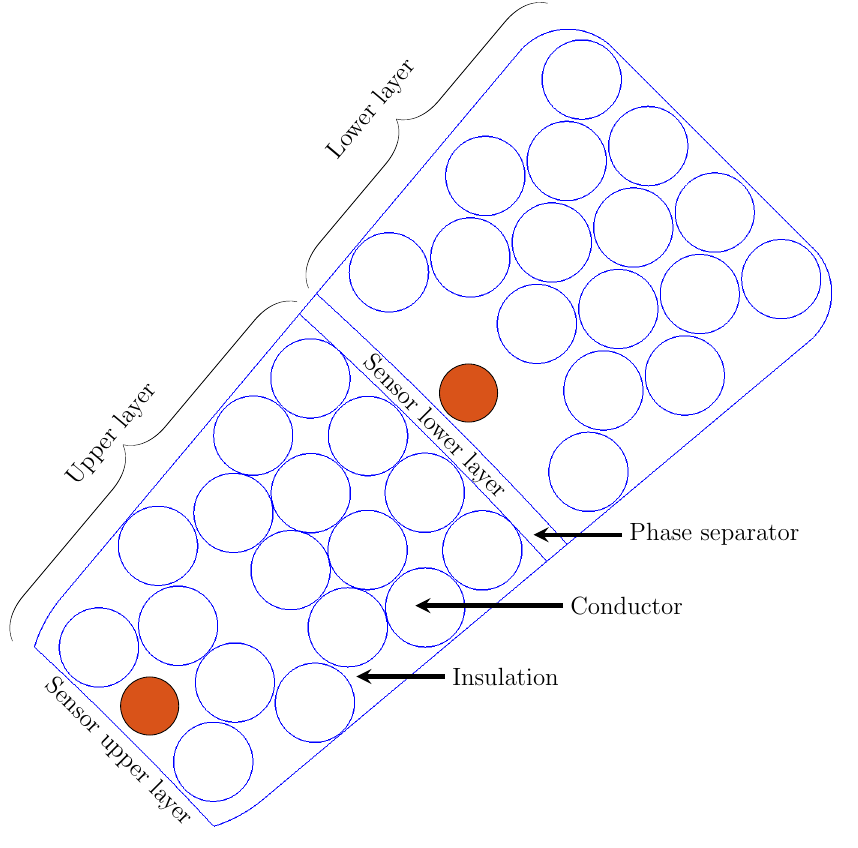}
	\caption{Position of the temperature sensors inside the stator slots at the axial center of the upper and lower winding.}
	\label{fig:4}
\end{figure}

\begin{table}
	\centering
	\caption{Data of the squirrel-cage induction machine with Y-connection}
	\resizebox{0.4\textwidth}{!}{
	\begin{tabular}{l|r@{\,}l}
		nominal data & value\\
		\hline
		nominal power & 3.7&kW \\
		nominal voltage RMS & 400&V \\
		nominal current RMS & 6.9&A \\
		nominal speed & 1430&rpm \\
		nominal frequency & 50&Hz \\
		number of pole pairs & 2& \\
		moment of inertia & 0.0195&kgm$^{2}$ \\
	\end{tabular}}
	\label{tab:1}
\end{table}
The machine is mounted on a test bench at the Electric Drives and Machines Laboratory of TU Graz \cite{Eickhoff_2020aa,Eickhoff_2021aa}. The measurement setup includes the machine under test and a connected heating device. The machine is fully equipped with a temperature measurement system, including rotor temperature detection via a telemetry system. Even though the machine is rated for air cooling, it is constructed with a cooling jacket to allow for various thermal experiments. Thermally insulating materials cover the cooling jacket, the end plates, and the stator flange (International Mounting B5) in order to reduce heat flows which are not monitored by the sensors. A flow meter and platinum resistance detectors (Pt100) are mounted at the inlet and outlet tube of the cooling jacket to monitor the cooling performance. The temperature sensors are positioned as follows: 18 temperature sensors are inserted in the middle of the lower layer and the upper layer of the stator slots.
The sensors of the lower and upper winding layer are located close to the slot separators and the wedges, respectively (see Fig.~\ref{fig:4}). Additionally, 12 sensors are positioned inside and on the surface of the winding overhang. The temperature sensors of the slots and the winding overhang are thin film resistance temperature detectors (2x10mm) according to DIN EN 60751 \cite{DIN_60751}. Temperature sensors of the same type but with smaller dimensions (2x4mm) monitor the temperature of the stator yoke, and are positioned at the axial center of the yoke. 
12 sensors are positioned in small holes drilled in the center of the stator teeth and the stator yoke (see Fig.~\ref{fig:3}). Finally, five sensors monitor thermal transients in the rotor and the shaft.

\section{Thermal Model}
\begin{figure}
    \centering
	\includegraphics[width=.8\linewidth]{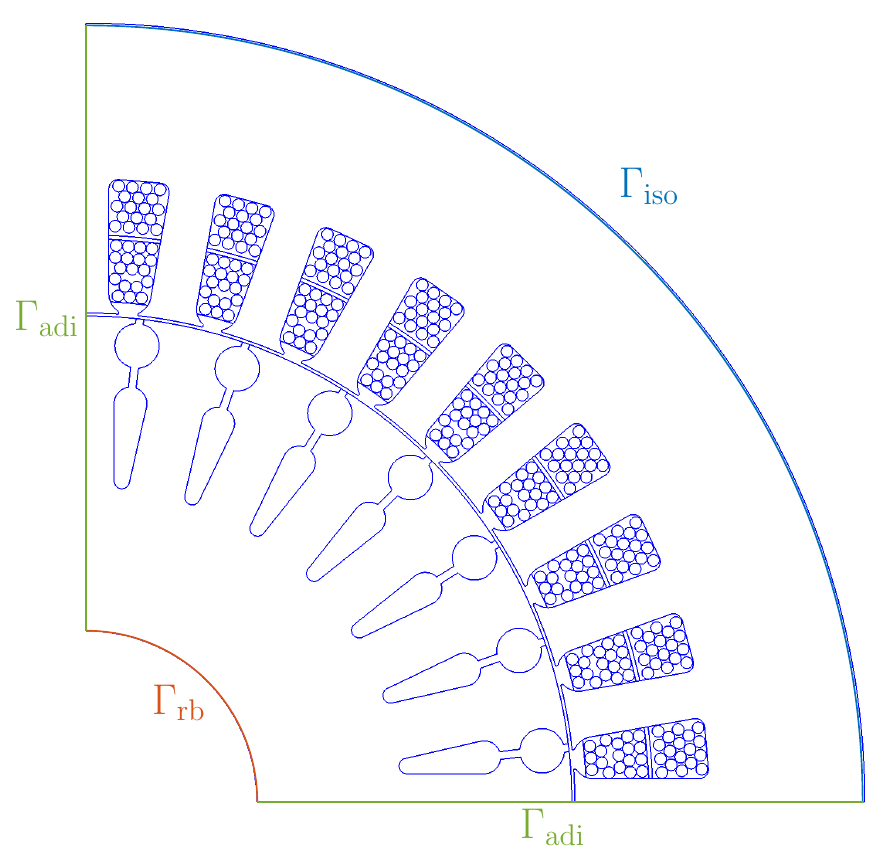}
    \caption{Geometry of the \ac{2d} model with the corresponding boundary conditions for the cooling jacket, the motor shaft and the rotation symmetry. The wires surrounded by homogeneous insulation material are shown in the slots.}
    \label{fig:5}
\end{figure}

The thermal behavior of the induction machine is modeled by the transient heat conduction equation \cite{Chemieingenieurwesen_2010aa} in a \ac{2d} \ac{fe} setting,
	\begin{align}
		{c_\text{v} \frac{\partial T}{\partial t} }- \nabla\cdot\left( {{\lambda}\nabla{T}}\right) &= \dot{q} \hspace{1cm} \text{in}\, V,\tag{1a}\label{eq:heatCond}\\
		T &= T_\text{iso}  \hspace{0.7cm} \text{on}\, \Gamma_\text{iso},\tag{1b}\label{eq:iso}\\
		-{\lambda}\frac{\partial T}{\partial \bm{n}} &= 0 \hspace{1cm} \text{on}\, \Gamma_\text{adi}\tag{1c},
		 \label{eq:adi}\\
		 {\lambda}\frac{\partial T}{\partial \bm{n}} + h(T-T_\text{iso})&=0 \hspace{1cm} \text{on}\, \Gamma_\text{rb}\tag{1d}.
		 \label{eq:conv}
	\end{align}
Herein, $T$ is the temperature, $c_{\text{v}}$ is the volumetric heat capacity, ${\lambda}$ is the thermal conductivity, $h$ is the heat-transfer coefficient, and $\dot{q}$ is heat flux density. The computational domain is denoted as $V$, its boundary as $\Gamma$ and the normal vector at the boundary as $\mathbf{n}$. 
The \ac{2d} modeling approach is motivated by the radial dominance of the temperature gradient. This predominance is attributed to two key factors: the presence of a cooling jacket and the anisotropic nature of the thermal conductivities due to laminating the stator and rotor yokes. Specifically, the radial thermal conductivities substantially exceed the axial thermal conductivities,  

Additionally, the following boundary conditions are enforced: An isothermal boundary condition \eqref{eq:iso} is prescribed on $\Gamma_\text{iso}$ to account for the uniform temperature distribution of the cooling jacket. Due to the inherent symmetry of the machine, modeling a single quadrant suffices for accurate representation. Consequently, an adiabatic boundary condition is applied along the symmetry lines $\Gamma_\text{adi}$.
The rotor shaft gives rise to axial heat transfer to the externally mounted equipment, the test bench, and the ambient air. This modest yet significant \ac{3d} heat transfer is taken into account by the Robin-type boundary condition \eqref{eq:conv}, representing a weighted blend of Dirichlet and Neumann boundary conditions \cite{DAngelo_2017aa}. It is applied along the shaft surface line $\Gamma_\text{rb}$. Figure~\ref{fig:5} shows the \ac{2d} model including the imposed boundary conditions.

\section{Simulation and Measurement}
\subsection{Fitting parameters}

\begin{figure}
	\centering
	\setlength{\fwidth}{0.4\textwidth}
	{\input{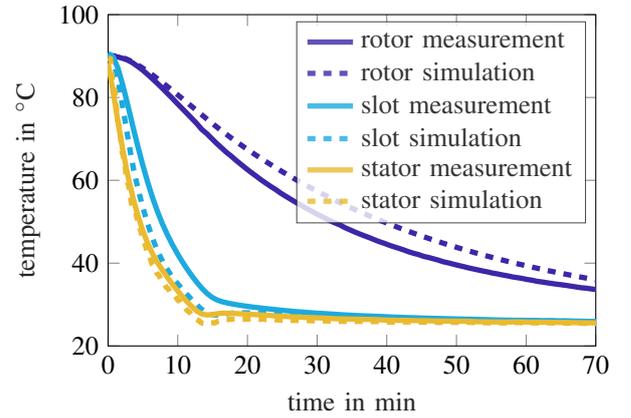}}
	\caption{Simulation results with the literature values of Table~\ref{tab:2} compared to the measured temperature decline over time for selected measurement positions in the model. The begin temperature is $93\,^\circ $C and the ambient temperature is $26\,^\circ $C. The overshoot at t=15\,min of the stator curve is due to the control of the cooling jacket.}
	\label{fig:6}
\end{figure}

\begin{figure}
	\centering
	\setlength{\fwidth}{0.4\textwidth}
	{\input{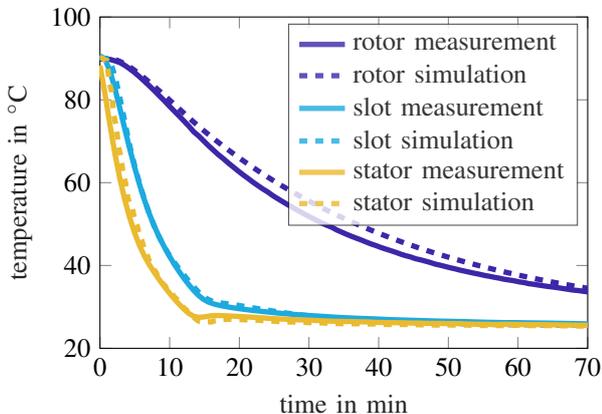}}
	\caption{Simulation results with the adjusted values of Table~\ref{tab:3} compared to the measured temperature decline over time for selected measurement positions in the model. The begin temperature is $93\,^\circ $C and the ambient temperature is $26\,^\circ $C.}
	\label{fig:7}
\end{figure}

\begin{table}
	\centering
	\caption{Thermal material parameters from literature }
	\begin{tabular}{c|c|c l}
		domain& parameter & value & unit \\
		\hline
		stator yoke & $\lambda_{\text{rho}}$    & 40 &[W/$^\circ$C/m] \\
		 & $\lambda_z$   &  2.5 &[W/$^\circ$C/m] \\
    		 & $c_\text{v}$ 	&  $3.925 \cdot10^6$ &[J/$^\circ$C/m$^3$] \\ \hline
		rotor yoke  &  $\lambda_{\text{rho}}$  	& 40& [W/$^\circ$C/m]  \\
		 & $\lambda_z$   &  2.5& [W/$^\circ$C/m] \\
		 &  $c_\text{v}$ 	&  $3.925 \cdot10^6$&[J/$^\circ$C/m$^3$]  \\ \hline
		conductor  &  $\lambda$  	&  398 &[W/$^\circ$C/m] \\
		  &  $c_\text{v}$ 	&  $3.435 \cdot10^6$ &[J/$^\circ$C/m$^3$] \\ \hline
		insulation  &  $\lambda$ 	& 0.7 &[W/$^\circ$C/m]  \\
		 & $c_\text{v}$  	& $7.905 \cdot10^6$ &[J/$^\circ$C/m$^3$] \\ \hline
		air gap  &  $\lambda$ 	& 0.026 &[W/$^\circ$C/m]  \\
		 & $c_\text{v}$  	& $1.210 \cdot10^3$ &[J/$^\circ$C/m$^3$] \\ \hline
		 shaft &  $\lambda$ 	& 59.6 &[W/$^\circ$C/m]  \\
& $c_\text{v}$  	& $3.777 \cdot10^6$ &[J/$^\circ$C/m$^3$] 
	\end{tabular}
	\label{tab:2}
\end{table}
Table~\ref{tab:2} provides a comprehensive list of the thermal material parameters obtained from manufacturer specifications and established literature. The simulation results in Fig.~\ref{fig:6} are obtained using those thermal material parameters. Clearly, the
temperature decline in the slot and in the stator is overestimated, while the temperature decline in the rotor is underestimated.
This leads to the assumption that certain factors introduce systematic uncertainties into the model, defying a priori determination.
These factors are:
\begin{itemize}
    \item Natural convection in the air gap: This effect is assumed to be minimal due to the small air gap. Notably, forced convection is absent as the rotor does not rotate.
    \item Axial heat transfer along the shaft: This effect is parameterized by $h$ and $\lambda$ in the Robin boundary condition \eqref{eq:conv}. 
    \item Anisotropic heat transfer: Anisotropic thermal conductivities of the stator yoke and rotor lead to an anisotropic heat transfer. 
    \item Lamination: This characteristic arises from the anisotropic nature of the thermal conductivities in the lamination. To appropriately account for this effect within a two-dimensional model, an effective thermal conductivity is introduced. 
\end{itemize}
These factors undergo a fitting process against the cool-down measurements of  Sec.~\ref{sec:Cooldown}, enabling a refinement of their initial values.
\begin{table}
	\centering
	\caption{Fitted thermal material parameters }
	\begin{tabular}{c|c|c l}
		domain& parameter & value & unit \\
		\hline
		stator yoke & $\lambda_{\text{eff}}$    & 24 &[W/$^\circ$C/m] \\\hline
		rotor yoke  &  $\lambda_{\text{eff}}$  	& 16& [W/$^\circ$C/m]  \\\hline
		air gap  &  $\lambda_{\text{eff}}$ 	& 0.052 &[W/$^\circ$C/m]  \\ \hline
		 shaft &  $\lambda_{\text{eff}}$ 	& 59.6 &[W/$^\circ$C/m]  \\
		  & $h$ & 0.235 & [W/$^\circ$C/m]
	\end{tabular}
	\label{tab:3}
\end{table}
Table~\ref{tab:3} lists the fitted parameters of the machine model and Fig.~\ref{fig:7} shows the improved temperature behavior. The obtained effective thermal conductivities of 16 W/$^\circ$C/m in the stator yoke and 24 W/$^\circ$C/m in the rotor are in good accordance to the thermal conductivity suggested in \cite{Wrobel_2015aa}.

The \ac{2d} \ac{fe} model of the machine is simulated and analyzed with the open source Python-based \ac{fe} framework \textit{Pyrit} \cite{Bundschuh_2023ab}. 

\subsection{Cooldown process scenario}\label{sec:Cooldown}
In a first step, the cooling characteristics of the induction machine are analyzed by conducting a series of exclusively thermal experiments, i.e. experiments without any electric excitation. In these experiments a defined temperature profile is prescribed by the cooling jacket: First, the machine is heated until a steady-state is reached across all machine domains. Then, the cooling jacket temperature is reduced to ambient conditions of $26\,^\circ$C and the cool-down process of the machine is recorded.

\begin{table}
	\centering
	\caption{Thermal time constants $\tau$ in (min)}
	\begin{tabular}{c|c|c|c|c}
	    initial & domain& measurement & simulation & rel. error \\
	    temperature &&&&\\
		
		\hline
		$93\,^\circ$C & slot   & 7.01 & 6.97 & 0.6\,\%\\
			          & stator yoke  & 4.94 & 4.88 & 1.6\,\%\\
			          		\hline
		$83\,^\circ$C & slot   & 6.81 & 6.81 & 0.02\,\%\\
			          & stator yoke  & 4.82 & 4.75 & 1.4\,\%\\
			          		\hline
		$73\,^\circ$C & slot   & 6.77 & 6.77 & 0.04\,\%\\
			          & stator yoke  & 4.78 & 4.70 & 1.5\,\%\\
			          		\hline
		$63\,^\circ$C & slot   & 7.24 & 7.18 & 0.6\,\%\\
			          & stator yoke  & 5.19 & 5.29 & 1.9\,\%\\
			          		\hline
		$53\,^\circ$C & slot   & 5.73 & 5.81 & 1.5\,\%\\
			          & stator yoke  & 3.98 & 3.90 & 2.0\,\%\\
			          		\hline
		$45\,^\circ$C & slot   & 6.17 & 6.21 & 0.7\,\%\\
			          & stator yoke & 4.27 & 4.33 & 1.4\,\%
	\end{tabular}
	\label{tab:4}
\end{table}
The cooling behavior inside the rotor, stator and slots starting from an initial temperature of 93$\,^\circ$C is shown in Fig.~\ref{fig:7}. The temperature decline in the rotor is significantly slower compared to the stator due to the low thermal conductivity of the air gap (see Tab.~\ref{tab:2}). The simulation results in the stator domain are in very good agreement with the measurement data. A slightly larger deviation can be seen in the rotor domain (see Fig.~\ref{fig:7}). This can be attributed to an axial heat transfer in $z$-direction from the rotor shaft to the attached torque-generator which is not considered by the \ac{2d} model. Additionally, while the stator sensors are drilled into the longitudinal center of the machine, the rotor sensors are mounted onto the rotor surface. This might lead to additional unmonitored parasitic effects, e.g. the influence of the surrounding air. 

To assess the validity of the simulation results, the relative error of the thermal time constants is computed. 
The thermal time constant is defined as the time required for a temperature drop of 63\,\% from the initial steady-state temperature. The relative error is computed as
\begin{equation}
    \varepsilon_{\text{rel}}=\frac{|\tau_{\text{meas}}-\tau_{\text{sim}}|}{\tau_{\text{meas}}}\,,\tag{2}
\end{equation}
where $\tau_{\text{meas}}$ and $\tau_{\text{sim}}$ are the thermal time constants corresponding to measurement and simulation, respectively.

\begin{figure}
	\centering
	\setlength{\fwidth}{0.4\textwidth}
	{\input{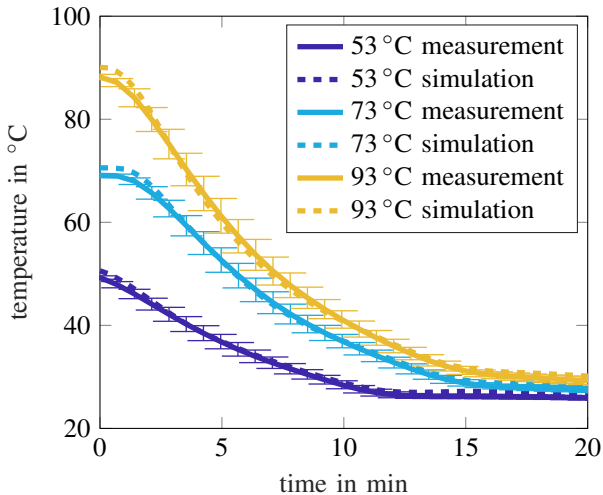}}
	\caption{Comparison between simulation and measurements for different steady-state target temperatures in the area of the slot. The error bars show the deviation between the measurements in all of the 36 slots.}
	\label{fig:8}
\end{figure}

Table~\ref{tab:4} compares the measured and simulated time constants for different steady-state target temperatures, $T_{\text{st}}$, ranging from 45$^{\circ}$C to 93$^{\circ}$C. To account for measurement errors, the mean value of all temperature sensors in the slots and the stator yoke, respectively, is considered. Figure~\ref{fig:8} shows the range of the temperature deviation between the sensors in the slot for three different steady-state target temperatures. Additionally, the resulting mean value of all temperature sensors and the simulation results are displayed. For the slots, the relative error lies within $0.02\,\%$ and $1.5\,\%$ and for the stator yoke within $1.4\,\%$ and $2.0\,\%$. This relative error is smaller than the measurement tolerance of the sensors, therefore the model is successfully validated for the cooldown scenario.

\subsection{Load cycle scenarios}
\begin{figure}
	\centering
	\includegraphics[width=\linewidth]{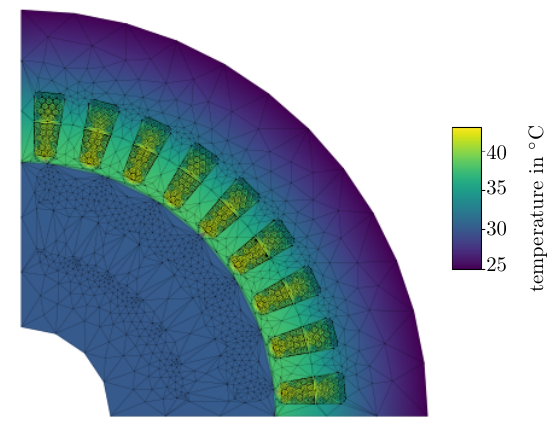}
	\caption{Steady-state temperature distribution in one quarter of the cross section of the machine for a power supply of 200\,W.}
	\label{fig:9}
\end{figure}

\begin{figure}
	\centering
	\setlength{\fwidth}{0.4\textwidth}
	{\input{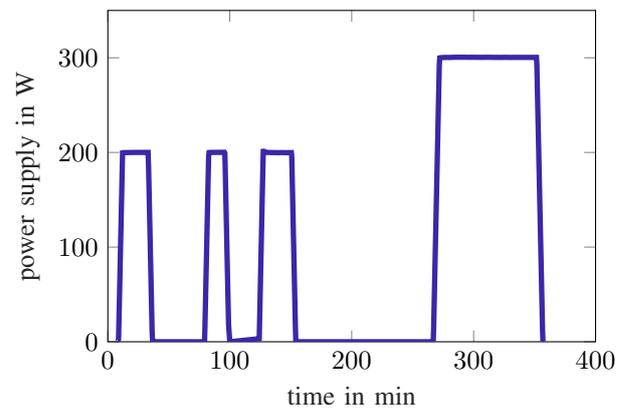}}
	\caption{The power of 200\,W and 300\,W are applied, until a steady temperature distribution in the machine in reached. Then the power supply is set to 0\,W in order to let the machine cool down again.}
	\label{fig:10}
\end{figure}

\begin{figure}
	\centering
	\setlength{\fwidth}{0.4\textwidth}
    \input{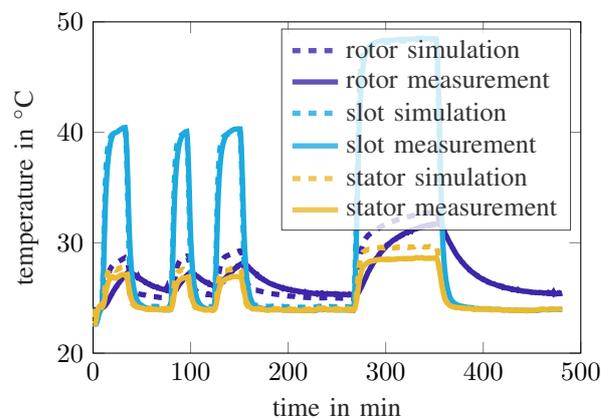}
	\caption{Load cycle scenario. In this case multiple measurements with a fixed ambient temperature are carried out.  The compared positions are in the rotor yoke, stator yoke and slot.}
	\label{fig:11}
\end{figure}

The model is now applied to study a second heating scenario. Several load cycles with varying power levels are performed and the resulting temperature distributions are evaluated. The machine is energized with all windings series connected supplied by a DC current such that no movement of the rotor takes place. The power supply is ramped up slowly at a rate of 1\,W/s such that the magnetic losses are negligible. Consequently, the supplied power is completely converted to heat by the Joule losses inside the stator windings. Additionally, the cooling jacket keeps a steady ambient temperature of 26\,$^\circ $C.
Figure~\ref{fig:10} shows the power supply during several load cycles with power levels of 200\,W and 300\,W, respectively. The resulting temperatures are shown in Fig.~\ref{fig:11}. Since the windings act as a primary heat source, the maximum temperature is located in the slot area (see Fig.~\ref{fig:9}) and therefore also close to the insulation system. The maximum temperatures in the slot area are $40\,^\circ$C and $48\,^\circ$C for the 200\,W and 300\,W load cycles, respectively.

\begin{figure}
	\centering
	\setlength{\fwidth}{0.4\textwidth}
	{\input{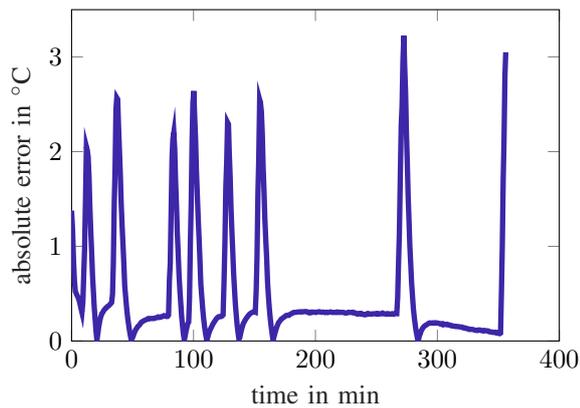}}
	\caption{Absolute error between measurement and simulation for the slot. The high peaks are due to the fast change in the temperature in the slot area, after switching of the power supply.}
	\label{fig:12}
\end{figure}

Comparing the experimental and simulation data, also for this scenario, a good agreement can be observed. The absolute error of the simulated temperature in the slot area over time is displayed in Fig.~\ref{fig:12}. The absolute error fluctuates between approximately $0.3\,^\circ$C during the plateau phases and $3\,^\circ$C during the heating and cooling phases. The larger error during the transition phases are due to the fact that the time instant when the power supply changes is only known up to one minute, leading to slight differences in the switching times of measurement and simulation. The general good agreement is considered as a further validation of the model.

\section{Conclusion}
A thermal stress analysis of the insulation system is a crucial part for designing a robust machine. Presently, the design process often overlooks the insulation system's reliability and longevity. This work presents a thermal model of a squirrel-cage induction machine. As the predominant temperature gradient is radial, a \ac{2d} \ac{fe} model is both accurate as well as computationally cheap. A series of transient thermal experiments is conducted and the results are used to assess the quality of the \ac{fe} model. The comparison between measurements and simulations shows a very good accordance. In conclusion, the \ac{fe} model is successfully validated. 

\section{Acknowledgment}
This work is funded by the Deutsche Forschungsgemeinschaft (DFG, German Research Foundation) – Project-ID 492661287 – TRR 361, the Athene Young Investigator Programme of TU Darmstadt and the Graduate School Computational Engineering at TU Darmstadt. We thank Greta Ruppert for the fruitful discussions.

\printbibliography

\begin{acronym}
	\acro{2d}[2D]{two-dimensional}
	\acro{3d}[3D]{three-dimensional}
	\acro{ac}[AC]{alternating current}
	\acro{dc}[DC]{direct current}
	\acro{dof}[DoF]{degree of freedom}
	\acroplural{dof}[DoFs]{degrees of freedom}
	\acro{em}[EM]{electromagnetic}
	\acro{epdm}[EPDM]{ethylene propylene diene monomer rubber}
	\acro{eqs}[EQS]{electroquasistatic}
	\acro{eqst}[EQST]{electroquasistatic-thermal}
	\acro{es}[ES]{electrostatic}
	\acro{fe}[FE]{finite element}
	\acro{fem}[FEM]{finite-element method}
	\acro{fgm}[FGM]{field grading material}
	\acro{hv}[HV]{high-voltage}
	\acro{hvac}[HVAC]{high-voltage alternating current}
	\acro{hvdc}[HVDC]{high-voltage direct current}
	\acro{lsr}[LSR]{liquid silicone rubber}
	\acro{sir}[SiR]{silicone rubber}
	\acro{mo}[MO]{metal oxide}
	\acro{pd}[PD]{partial discharges}
	\acro{pde}[PDE]{partial differential equation}
	\acro{pea}[PEA]{pulsed-electro-acoustic method}
	\acroplural{pde}[PDEs]{partial differential equations}
	\acro{pm}[PM]{person month}
	\acroplural{pm}[PMs]{person months}
	\acro{qoi}[QoI]{quantity of interest}
	\acroplural{qoi}[QoIs]{quantities of interest}
	\acro{rdm}[RDM]{research data management}
	\acro{rms}[rms]{root mean square}
	\acro{tna}[TNA]{Transient Network Analysis}
	\acro{tuda}[TUDa]{TU~Darmstadt}
	\acro{tum}[TUM]{TU~Mün\-chen}
	\acro{uq}[UQ]{uncertainty quantification}
	\acro{wp}[WP]{work package}
	\acroplural{wp}[WPs]{work packages}
	\acro{xlpe}[XLPE]{cross-linked polyethylene}
	\acro{zno}[ZnO]{zinc oxide}
	\acro{CI}[CI]{continuous integration}
\end{acronym}

\end{document}